# Kinetic Expression for Optimal Catalyst Electronic Configuration: The Case of Ammonia Decomposition

Nigora Turaeva[1]*, Rebecca Fushimi[2], Gregory Yablonsky[3]

[1] Webster University, Saint Louis, MO 63119
[2] Idaho National Laboratory, Idaho Falls, ID 83415
[3] Washington University in Saint Louis, Saint Louis, MO 63103

* Corresponding author, nigoraturaeva82@webster.edu

**Abstract**

A new steady-state kinetic model of ammonia decomposition is presented and analyzed regarding the electronic properties of metal catalysts. The model is based on the classical Temkin-Ertl mechanism and modified in accordance with Wolkenstein's electronic theory by implementing participation of free electrons of the catalyst to change the chemical nature of adsorbed species. Wolkenstein's original theory only applied to semiconductors but by including the *d*-band model for splitting of adsorbate molecular orbitals into bonding/antibonding states, the electronic theory can be extended to metals. The relative population of charged versus neutral adsorbates is a function of the Fermi level of the catalyst and the *d*-band splitting of the adsorbate. Moreover, charged and neutral adsorbates will present different reactivity and the overall reaction rate can be described as a function of the Fermi level. For both simplified and full reaction mechanisms, including electronic steps, we present a steady-state rate equation where the dependence on the Fermi level of the metal creates a volcano-shaped dependence. According to the kinetic model, an increasing Fermi level of the catalyst, that approaching the antibonding state with adsorbed nitrogen molecules, will increase the fraction of neutral nitrogen molecules and enhance their the desorption. Concurrently, strong chemisorption of ammonia molecules proceeds easily through participation of *additional free* catalyst electrons in the adsorbate bond. As a result, the reaction rate is enhanced and reaches its maximum value. A further increasing Fermi level of the catalyst that approaches the antibonding state with ammonia molecules will result in a smaller fraction of negatively charged ammonia molecules and less dehydrogenation. Concurrently, the desorption of neutral nitrogen molecules occurs without impairment. As a result, the reaction rate decreases. The detailed kinetic model is compared to recent experimental measurements of ammonia decomposition on iron, cobalt and CoFe bimetallic catalysts. This result agrees with the classical qualitative Sabatier statement as well as the quantitative theoretical results of the optimal binding energy present by Nørskov and coworkers. In difference from the classical volcano-curves, considering a single reactant, the volcano-shaped reaction rate of the presented model is obtained via the interplay of both reactant adsorption and product desorption, which is achieved by considering the role of electronic subsystems.

1. **Introduction**

It is widely recognized that French chemist Paul Sabatier was the first to express the importance of optimal adsorption where the interaction between the catalyst and adsorbed species should be neither too strong nor too weak. Sabatier's principle is manifest in the so-called 'volcano-plots', introduced by Balandin [1], which are now widely used to describe an optimal reaction rate as a function of adsorbate/catalyst properties such as the heat of adsorption. In many theoretical papers, the justification for the volcano relationship arises from the linear Brønstead- or Bell-Evans-Polanyi (BEP) equation describing the dependence of the apparent activation energy on the global reaction enthalpy. Additionally, the volcano relationship has been described as an interplay between the opposing processes of adsorption and desorption. Volcano relationships are generally expressed relative to *molecular properties*, e.g. binding energy and heat of adsorption, and based on the breaking/formation of bonds in a body of several atoms.

Quantum mechanical calculations have greatly enabled the study of these dependences and the well-established *d*-band theory [2] takes the molecular relationship a step further, to include *electronic properties* when adsorbate molecular orbitals are split into bonding and antibonding states via hybridization with the narrow *d*-band electronic states of a metal catalyst. The Fermi level of the metal determines to what extent these hybrid orbitals are filled and therefore the strength of the metal-adsorbate bond as well. For example, if the Fermi level falls within the antibonding states then the metal-adsorbate bond will be weakened. Thus, the *difference* between the Fermi level and the hybridized adsorbate bonding/antibonding states is a crucial factor for describing the bond stability.

At the same time, the Fermi level of a metal indicates the availability of *free electrons* to localize through metal-adsorbate bonding leading to the creation of charged surface species. Herein we derive a new kinetic expression that includes electronic factors for predicting a volcano-type dependence of the reaction rate as a function of the Fermi level of a metal catalyst. Starting with the conventional elementary reaction mechanism for ammonia decomposition, we add electronic steps based on the participation of additional free valence electrons from the metal through the covalently bonded molecular orbitals of adsorbed species. Following adsorption, the fraction of charged species can be predicted based on the Fermi level of the catalyst. The same adsorbed species when either charged or neutral (*e.g.* $ZNH_3^-$ vs. $ZNH_3^0$) will have different reactive properties; hence the concentration of surface intermediates, the reaction kinetics and reaction pathways are all sensitive to the Fermi level. Thus, instead of the direct filling of bonding/antibonding adsorbate orbitals, we indirectly use *d*-band theory to understand how the fraction of charged species for both reactant and products impact the rate determining step of the reaction mechanism.

The derived kinetic expression describes a counterbalance between the rate-limiting steps of ammonia adsorption and nitrogen release from an electronic perspective: At small values of the Fermi level, a fraction of neutral nitrogen molecules is small compared to strong chemisorbed ones, and the reaction rate is limited by desorption of nitrogen molecules while a fraction of negatively charged ammonia surface species is high, and chemisorption of ammonia is strong enough for dehydrogenation. At large values of the Fermi level, the reaction rate is limited by adsorption of ammonia molecules, while the desorption of nitrogen molecules occurs without impairment. Such effects will be described in relationship to the dominating surface species observed in previously reported transient kinetic experiments of ammonia decomposition on cobalt, iron and bimetallic CoFe materials. The prediction of the rate volcano-dependence is also compared to reported steady-state characterization of ammonia decomposition on the same materials. The interplay of both electronic and kinetic factors provides a new perspective for predicting optimal catalyst composition and volcano-type dependences which have in the past typically been described based primarily on cohesive forces. Moreover, from the rate maximum of our kinetic expression, an optimal Fermi level can be described as a function of the arithmetic mean of the antibonding LUMOs of the adsorbed reactant (ammonia) and the product (nitrogen) in addition to a polynomial of kinetic constants.

## 2. Traditional Kinetic Model of Ammonia Decomposition

Decomposition of ammonia is a reverse reaction of the Haber-Bosch synthesis process which has been one of the most intensively investigated reactions for more than 100 years. Aside from its role in producing synthetic fertilizers, ammonia has a remarkable capacity for chemical storage of hydrogen. A significant literature exists for ammonia decomposition studies that describe how kinetic dependencies are influenced by different catalysts (supporting materials and promoters) as well as reaction parameters (temperature, concentration, etc.) [3-10]. In the literature, two possible rate limiting steps are discussed which generally depend on the catalyst metal component and process conditions: either the ammonia adsorption step or the nitrogen desorption step [9, 11-13]. Since at low temperatures nitrogen desorption was found to be the rate limiting step irrespective of the catalyst [13], the metal-N-binding energy was chosen as a key parameter in the design of catalysts for ammonia decomposition [4, 8]. From experimental data, reproduced in Figure 1, it was shown by Boisen *et al.* that for different monometallic metals the relationship between the ammonia decomposition rate and the nitrogen binding energy demonstrates a volcano shape; with the optimum corresponding to ruthenium metal [8]. This volcano-type dependence reflects two opposing ways for increasing activity of a single metal: either increasing the surface N-binding energy for greater surface N-coverage from ammonia conversion or weakening the surface N-binding energy to reduce the barrier for $N_2$ formation.

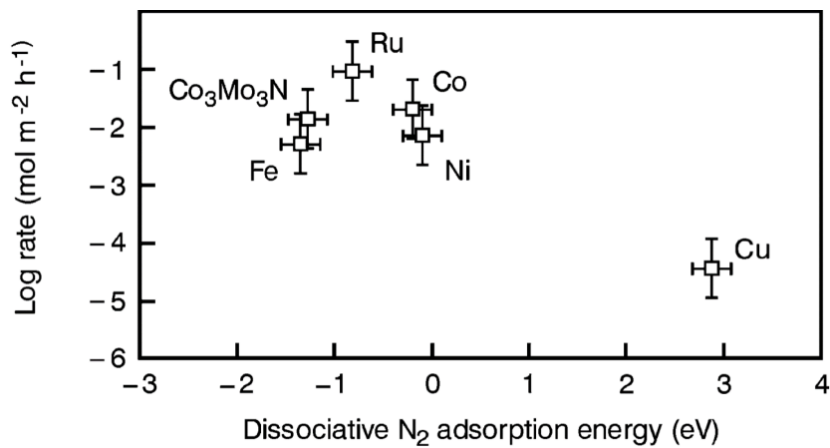

*Figure 1: Experimental rate of ammonia decomposition over various catalysts at 773 K, 1 bar, 3:1 $H_2/N_2$, and 20% $NH_3$ as a function of the reaction energy of dissociative $N_2$ adsorption. Reproduced with permission from Elsevier, Boisen* et al. *[8].*

In the synthesis of ammonia, a six-step mechanism was proposed by Temkin [14, 15] and the same steps in reverse were used by Ertl to describe ammonia decomposition [16, 17]. As described in Table 1, this reaction is initiated by the adsorption of ammonia and then proceeds with successive dehydrogenation of $NH_3$ releasing surface hydrogen atoms that combine to form molecular hydrogen. Once dehydrogenation is complete, nitrogen atoms may combine to form molecular nitrogen in the final step of the mechanism. In Table 1, *Z* is an active site of the catalyst and $ZNH_3$, $ZNH_2$, $ZNH$, $ZN$, and $ZH$ are different surface intermediates.

*Table 1: Conventional six-step mechanism for ammonia decomposition.*

| | |
|---|---|
| $NH_3 + Z \leftrightarrow ZNH_3$ | (1-1) |
| $ZNH_3 + Z \leftrightarrow ZNH_2 + ZH$ | (1-2) |
| $ZNH_2 + Z \leftrightarrow ZNH + ZH$ | (1-3) |
| $ZNH + Z \leftrightarrow ZN + ZH$ | (1-4) |
| $2ZH \leftrightarrow H_2 + 2Z$ | (1-5) |
| $2ZN \leftrightarrow N_2 + 2Z$ | (1-6) |

## 3. Electronic Theory Prediction of the Distribution of Charged Surface Species

From the traditional six-step mechanism for ammonia decomposition, Table 1, our design is to incorporate additional steps that include the participation of charged surface species created through the contribution of free metal electrons to adsorbate bonds. In early work, by supposing that surface electrons localize around an adsorbate and contribute to bonding in the same way as electrons in a molecule, Eley extended Pauling's formalism on electronegativity using the work function to describe the participation of conduction electrons in bonding [18], see also discussion by R.I. Masel [19]. More directly, the participation of free charge carriers (electrons and holes) in adsorbate bonding, and the derivation of kinetic effects, was described in the electron theory of catalysis on *semiconductor* materials by Th. Wolkenstein[1] in the 1960s [20, 21]. This theory was, explicitly, not applicable to metals, but, as we aim to demonstrate here, the addition of the more recently developed *d*-band theory can provide what was missing to extend this body of work into the domain of metal catalysts.

Wolkenstein described adsorption as creating a structural defect and disturbance in the electronic state of a pristine surface and hence a *localization center* for free charge carriers. Upon adsorption, Wolkenstein described the formation of acceptor and donor levels *within the bandgap of the semiconductor* where electronic transitions can take place with the valence and conduction band, *viz.* Figure 2. In this electronic theory, chemisorption can take on different forms: strong chemisorption takes place with the additional participation of free charge carriers (creating a charged adsorbate) and weak chemisorption does not (the adsorbate remains electrically neutral).[2] While electrons of the catalyst may form strong acceptor bonds with the adsorbed molecule, positive charges of the surface (holes) may form strong donor bonds with the adsorbate.

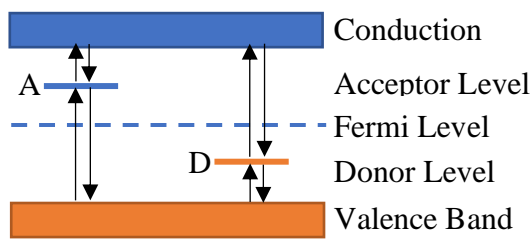

*Figure 2: Scheme for positions of acceptor, A, and donor, D, levels of an adsorbed molecule with electronic transitions from the valence and conduction bands of a semiconductor. Adapted from reference [21], Th. Wolkenstein.*

Of course, with metals, there are no forbidden states between the valence and conduction band, the valence band is typically partially filled (at least for the metals commonly used in catalysis) and the creation of discrete acceptor/donor levels should have no benefit for localizing electrons. If, however, we introduce the concepts derived in *d*-band theory, whereby the narrow *d*-band states of a transition metal induce splitting of adsorbate molecular orbitals into bonding and antibonding states, then we have an avenue for localizing free charge carriers as proposed by Wolkenstein and can add to our kinetic mechanism the participation of charged surface adsorbates. According to *d*-band theory, a difference in transition metal properties can be attributed to the width of the *d*-band states (a narrower band results in stronger coupling and a greater energy difference between bonding/antibonding hybrid orbitals) as well as the location of the Fermi level. The location of the Fermi level[3] characterizes the degree to which the antibonding state is occupied and hence the strength of the bond between the adsorbate and the metal surface. Hammer and Nørskov described the splitting of states into bonding and antibonding as 'strong chemisorption' whereas a simple broadening of the states, through interaction with the *s*-band for example, was referred to as 'weak chemisorption' [2]; this should be clearly distinguished from Wolkenstein's definition which is based on the participation of free charge carriers in the adsorbate bond.

---

[1] a.k.a. F.F. Volkenstein
[2] In both chemistry and physics the notion of strong and weak bonding has acquired many different definitions that should not be confounded with Wolkenstein's definition.

For reference, a conventional scheme for *d*-band splitting of molecular orbitals is depicted in Figure 3, we only indicate the peak energy states in the density of states distribution for simplicity but more rigorously, a complex distribution would need to be considered. By taking into account the experimental values of the HOMO (highest occupied molecular orbital, -10.82 eV) and the LUMO (lowest unoccupied molecular orbital, -5.10 eV) of ammonia molecules [22], we can arrange the scheme of energy positions in Figure 3. Approximate energy positions of bonding and antibonding molecular orbitals formed between the *d*-band center and the LUMO and HOMO of ammonia are presented for illustrative purposes and will vary with the identity of the metal. Note that an experimental value for the LUMO of a nitrogen molecule has not been reported and its theoretical value has a broad range from -3 to -8 eV [23], therefore the position of the bonding/antibonding states of the nitrogen molecule on the catalyst surface is approximate.

---

[3] In solid state physics, the Fermi level is a hypothetical energy level where, at thermodynamic equilibrium, the probability of finding an electron would be ½; it includes potential energy and is often referred to as the electrochemical potential. The term should not be confused with the Fermi Energy which is more often encountered in quantum mechanics and represents an energy *difference*, usually corresponding to kinetic energy, at absolute zero.

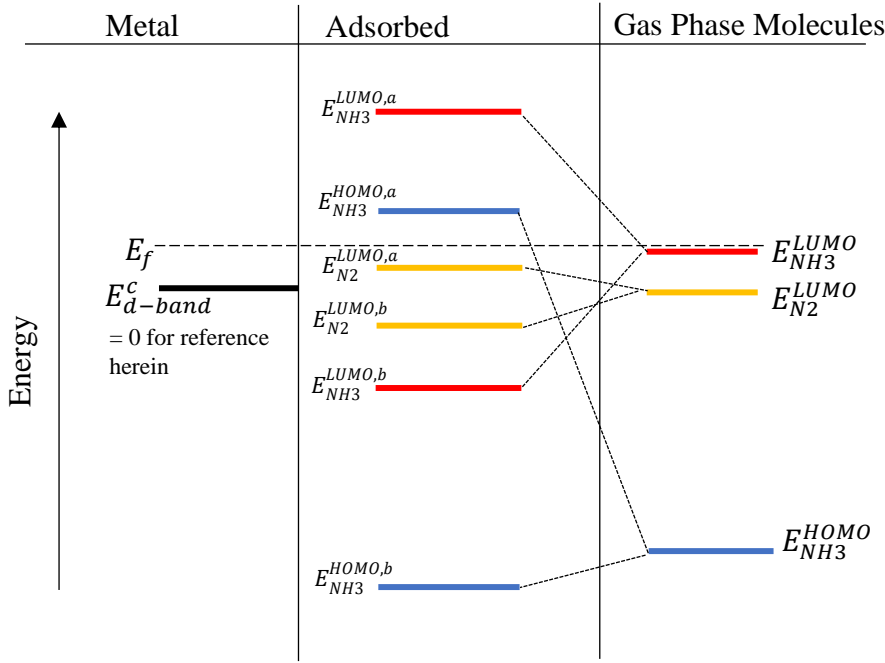

*Figure 3: Scheme for energy levels of a metal catalysts where $E_f$ is the Fermi level, $E_{cd\text{-band}}$ is the center of a narrow distribution of metal d-band states. The HOMO and LUMO states of an ammonia molecules are, $E_{NH3}^{LUMO}$ and $E_{NH3}^{HOMO}$, while only the nitrogen LUMO, $E_{N2}^{LUMO}$, is shown for clarity. These states are split into bonding and antibonding states upon adsorption via hybridization with the narrow d-band states of the metal. Approximate states are configured for illustrative purposes. The bonding state of the adsorbed nitrogen molecule is indicated for reference, $E_{N2}^{LUMO,b}$. We have set $E_d^c = 0$ as a reference point in our discussion.*

The positioning of the antibonding LUMO of the adsorbed species relative to the catalyst Fermi level will predict whether a free electron from the metal is likely be accepted, leading to a charged adsorbate. In this manner, Wolkenstein used the similar acceptor level, created upon adsorption and located within the semiconductor bandgap, to accept electrons from the valence or conduction bands, Figure 2. Such an event leads to the formation of a 'strong acceptor bond' and negatively charged surface species. Electronic transitions to and from the acceptor level correspond to transformation between strong to weak chemisorbed states and these fast electronic transitions quickly establish equilibrium. The fraction of charged chemisorbed species represented by the number density of different species,

$$\eta^0 = \frac{N^0}{N}, \quad \eta^- = \frac{N^-}{N}, \tag{1}$$

Where $N$ represents the total number of adsorbates and $\eta_0 + \eta_- = 1$. Using the Fermi-Dirac distribution we can predict the fraction of charged chemisorbed species according to the electrochemical potential of the metal and the adsorbed species:

$$\frac{\eta^0}{\eta^-} = e^{\frac{E_f - E_{ads}^{LUMO,a}}{kT}}, \tag{2}$$

where $E_f$ represents the Fermi level of the catalyst surface and $E_{ads}^{LUMO,a}$ represents the energy of antibonding LUMO of the adsorbate. Note that the energies are used as positive values with respect to a reference point set at the d-band center ($E_d^c = 0$).

The Fermi level of the metal determines the availability of free charge carriers in the solid. If free charge carriers from the metal can localize through the metal–adsorbate bond, then the resulting charged surface species will have different reactive properties compared to the same neutral species. The ratio of charged to neutral species is a function of the Fermi level of the metal and the antibonding LUMO of the adsorbates. The latter changes according the center of the *d*-band and to observe the dependence of the fractions of intermediates upon the Fermi level, we fix a reference point at the *d*-band center ($E_d^c = 0$). As the Fermi level moves down towards the *d*-band center, more antibonding adsorbate-metal *d*-states become empty and the ratio of fractions in Eqn. (2) decreases. Changes in the dominating surface intermediates and reaction pathways could be predicted based on this ratio. Hereafter, we will interpret the consequences of different positioning of the catalyst Fermi level relative to antibonding LUMOs of the adsorbed reactant and product.

## 4. Electronic Model of Ammonia Decomposition

In order to understand the different activity of metal catalyst we add to the general mechanism for ammonia decomposition, Table 1, the steps found in Table 2 that include the participation of electrons. Step (2-1) is simply the adsorption of ammonia in the neutral state. For any adsorbed species we can write an equilibrium reaction between neutral and charged states, *e.g.* step (2-1*). Recall that this equilibrium depends on the Fermi level of the metal and the LUMO of the adsorbed species, Eqn. (2). Since electronic processes are much faster than atomic rebuildings, we will assume that electronic equilibrium is established among the surface species. Both the charged and neutral species can proceed along the dehydrogenation pathway (steps (2-2) – (2-4)) but for simplicity we will only write such steps for the charged species. Moreover, based on established literature, we only consider either the ammonia adsorption or nitrogen desorption steps to be rate limiting [9, 11-13, 24], other electronic steps will be irrelevant for our analysis. Likewise, the recombination and desorption of gas phase hydrogen, as described earlier in step (1-5), is not considered rate limiting, does not influence the electronic analysis and is hence disregarded here. Step (2-5) describes the reversible formation of molecular nitrogen. As written, this step is not elementary and implicit in the scheme is the fast donation of one electron back to the metal and the combination of one neutral and one charged nitrogen species to form the charged nitrogen molecule. Since the metal has abundant electrons, rigorous accounting of electrons should not be necessary. Step (2-5*) represents the equilibrium reaction between the charged and neutral molecular nitrogen species. Desorption can only occur from a neutral state thus, after the catalyst has accepted the electron back from the N₂ adsorbate in the forward reaction of step (2-5*), desorption can proceed according to step (2-6). For simplicity we will not include reverse reactions of gas-phase products.

*Table 2: Electronic model for ammonia decomposition.*

| | |
|---|---|
| $NH_3 + Z \leftrightarrow ZNH_3^0$ | (2-1) |
| $ZNH_3^0 \leftrightarrow ZNH_3^-$ | (2-1*) |
| $ZNH_3^- + Z \rightarrow ZNH_2^- + ZH$ | (2-2) |
| $ZNH_2^- + Z \rightarrow ZNH^- + ZH$ | (2-3) |
| $ZNH^- + Z \to ZN^- + ZH$ | (2-4) |
| $2ZN^- \leftrightarrow ZN_2^- + Z$ | (2-5) |
| $ZN_2^- \leftrightarrow ZN_2^0$ | (2-5*) |
| $ZN_2^0 \to N_2 + Z$ | (2-6) |

Following the adsorption step, desorption of ammonia may proceed from the neutral state ($ZNH_3^0$) or the bonding LUMO of the adsorbed species can localize an electron from the metal valence states to form $ZNH_3^-$. Whereas surface adsorbates must desorb from the neutral state ($ZNH_3^0$ and $ZN_2^0$), surface reactions in charged states ($ZNH_3^-$, $ZNH_2^-$, $ZNH^-$, $ZN^-$ and $ZN_2^-$) may proceed differently from the neutral state. The ratios of adsorbed species in different charged states of chemisorption, $\eta$, can be presented according to Eqn. (1) as:

$$\eta_{NH3}^0 = \frac{N_{NH3}^0}{N_{NH3}}; \; \eta_{NH3}^- = \frac{N_{NH3}^-}{N_{NH3}}; \; \eta_{N2}^- = \frac{N_{N2}^-}{N_{N2}}; \; \eta_{N2}^0 = \frac{N_{N2}^0}{N_{N2}} \tag{3}$$

where $N_{NH3} = N_{NH3}^- + N_{NH3}^0$ and $N_{N2} = N_{N2}^- + N_{N2}^0$.

The total number of surface intermediates ($ZNH_3$, $ZNH_2$, $ZNH$, $ZN$, $ZN_2$), might be referred to as 'Langmuir intermediates' which exist under quasi-steady state conditions. This term is used to distinguish said intermediates from the charged and neutral adsorbates which might be termed 'Wolkenstein intermediates' and exist under equilibrium conditions. Based on these intermediates, we will make both linear and non-linear analysis of the reaction kinetics using different approaches for the solution within a reduced mechanism. We will show that in both cases we can find criteria at which the volcano shaped dependence of the reaction rate on the Fermi level is readily apparent. This dependence is then compared with experimental results obtained for different catalysts.

## 5. Reduced Mechanism: Linear Analysis

Let us consider the following reduced mechanism of ammonia decomposition. To provide a rigorous solution to the mechanism, we use a linearized form for irreversible surface reaction steps of intermediates for mathematical convenience, Table 3. In this mechanism, steps (3-1) and (3-1*) are the same adsorption step as previous. Steps (3-2) and (3-3) represent sequential dehydrogenation. Step (3-4) is the association of nitrogen atoms, Step (3-4*) describes the transition of the adsorbed $N_2$ species between charged and neutral. Step (3-5) is the same nitrogen desorption step described previously.

*Table 3: Electronic model for the reduced mechanism of ammonia decomposition.*

| | |
|---|---|
| $NH_3 + Z \leftrightarrow ZNH_3^0$ | (3-1) |
| $ZNH_3^0 \leftrightarrow ZNH_3^-$ | (3-1*) |
| $ZNH_3^- \to ZNH^- + H_2$ | (3-2) |
| $2ZNH^- \to 2ZN^- + H_2$ | (3-3) |
| $2ZN^- \leftrightarrow ZN_2^- + Z$ | (3-4) |
| $ZN_2^- \leftrightarrow ZN_2^0$ | (3-4*) |
| $ZN_2^0 \rightarrow N_2 + Z$ | (3-5) |

The reduced mechanism of ammonia decomposition presented in Table 3 is simpler than the previous one (Table 2): seven steps instead of eight; seven intermediates, including electronic states ($Z, ZNH_3^0, ZNH_3^-, ZNH^-, ZN^-, ZN_2^-, ZN_2^0$) instead of nine ($Z, ZNH_3^0, ZNH_3^-, ZNH_2^-, ZNH^-, ZN^-, ZN_2^-, ZN_2^0, ZH$); the concentration of $ZNH_2^-$ is neglected; the step of hydrogen desorption is considered fast and incorporated to steps of *ZNH3* and *ZNH*. What is important that this mechanism is much less non-linear than the mechanism presented in Table 2. Instead of three steps, in which different intermediates react with Z (steps 2-2, 2-3, 2-4), we have only two steps (steps 3-2, 3-3). Five steps of this mechanism (steps 3-1, 3-1*, 3-2, 3-4*, 3-5) are linear regarding the intermediates since only one molecule of the intermediate participates in the reaction. Therefore, this mechanism is more suitable for the application of the linear theory of steady-state complex reactions to the analysis of its kinetic model [25].

For derivation of steady-state rate equations, we can use the graph theory [25] or a special equation for 1/R$_{steady-state}$, which was recently developed to distinguish the kinetic and thermodynamic constituents [26]. R is the steady-state reaction rate of this single route reaction.
According to this equation,

$$\frac{1}{R} = \sum_{i=1}^{n} \frac{1}{\tilde{k}_i}\left[1 + \sum_{j=1}^{n} \prod_{j} \frac{1}{\tilde{K}_j}\right] \quad (4)$$

Here $\tilde{k}_i$ is the apparent kinetic constant of the forward *i*th step which may include the concentration of gaseous species as a parameter:

$\tilde{k}_1 = k_1 C_{NH3}$; $\tilde{k}_{1*} = k_{1*}$; $\tilde{k}_2 = k_2$; $\tilde{k}_3 = 2k_3$; $\tilde{k}_4 = 2k_4$; $\tilde{k}_{4*} = k_{4*}$; $\tilde{k}_5 = k_5$.

Here $k_1, k_{1*}, k_2, k_3, k_4, k_{4*}, k_5$ are the kinetic coefficients of corresponding forward reactions. $\tilde{K}_j$ is the apparent equilibrium constant of *j*th step, which may include the concentration of gaseous species as a parameter:

$\tilde{K}_1 = K_1 C_{NH3}$; $\tilde{K}_{1*} = K_{1*}$; $\tilde{K}_4 = K_4$; $\tilde{K}_{4*} = K_{4*}$,

where $K_1, K_{1*}, K_4, K_{4*}$ are equilibrium constants of corresponding steps. Steps 3-2, 3-3, 3-5 are considered irreversible.

In Eqn. (4), every term corresponds to the *i*th step. It is a product of two characteristics:

(1) $\frac{1}{\tilde{k}_i}$ represents a kinetic function of $i$th step;

(2) $\left[1+\sum_{j=1}^{n}\prod\frac{1}{\tilde{K}_j}\right]$ represents a thermodynamic function. It is obtained under the condition that other (n-1) steps are considered under the equilibrium.

In two cases, this equation is significantly simplified:

(1) some steps are fast, and the terms corresponding to these steps disappear;
(2) some steps are irreversible, and some terms corresponding to the thermodynamic factors disappear as well.

For the reduced mechanism of ammonia decomposition, Eqn.

(4) is rewritten as:

$$\frac{1}{R} = \frac{1}{k_1 C_{NH3}} + \frac{1}{k_{1*}}\left(1+\frac{1}{K_1 C_{NH3}}\right) + \frac{1}{k_2}\left(1+\frac{1}{K_{1*}}+\frac{1}{K_{1*}K_1 C_{NH3}}\right) + \frac{1}{2k_3} + \frac{1}{2k_4} + \frac{1}{k_{4*}}\left(1+\frac{1}{K_4}\right) + \frac{1}{k_5}\left(1+\frac{1}{K_{4*}}+\frac{1}{K_{4*}K_4}\right) \quad (5)$$

We can rewrite Eqn. (5) in the following way:

$$\frac{1}{R} = \left(\frac{1}{k_1 C_{NH3}}+\frac{1}{k_2}+\frac{1}{2k_3}+\frac{1}{2k_4}+\frac{1}{k_5}\right) + \frac{1}{k_2}\left(1+\frac{1}{K_1 C_{NH3}}\right)\frac{1}{K_{1*}} + \frac{1}{k_5}\left(1+\frac{1}{K_4}\right)\frac{1}{K_{4*}} \quad (6)$$

In Eqn. (6) the first five terms do not depend on the electronic factors, two last terms contain the equilibrium constants of electronic steps, $K_{1*}$ and $K_{4*}$. Kinetic coefficients of electronic steps 1* and 4* are absent as they are fast, however the equilibrium constants of these steps are present and provide support for Wolkenstein`s theory.

For different chemisorption states, equilibrium fractions of adsorbed species depend upon the energy difference between the Fermi level of the catalyst and the antibonding LUMOs of the adsorbates. Then, as in Eqns. (1) and (2) we can write:

$$\frac{\eta_{NH3}^0}{\eta_{NH3}^-} = \frac{1}{K_{1*}} = e^{\frac{E_f - E_{NH3}^{LUMO,a}}{kT}}; \quad \frac{\eta_{N2}^-}{\eta_{N2}^0} = \frac{1}{K_{4*}} = e^{-\frac{E_f - E_{N2}^{LUMO,a}}{kT}} \quad (7)$$

and Eqn. (6) becomes:

$$\frac{1}{R} = A(k) + \frac{1}{k_2}\left(1+\frac{1}{K_1 C_{NH3}}\right)e^{\frac{E_f - E_{NH3}^{LUMO,a}}{kT}} + \frac{1}{k_5}\left(1+\frac{1}{K_4}\right)e^{-\frac{E_f - E_{N2}^{LUMO,a}}{kT}} \quad (8)$$

Here $A(k) = \frac{1}{k_1 C_{NH3}}+\frac{1}{k_2}+\frac{1}{2k_3}+\frac{1}{2k_4}+\frac{1}{k_5}$ does not depend upon the electronic steps. It is seen from Eqn. (8), that the reaction rate represents a volcano-shape dependence on the Fermi level. The volcano-shape arises as a function of the Fermi level of different metals due to the competition between exponential terms describing the energy separation of the Fermi level of the catalyst and the antibonding LUMOs of the adsorbed ammonia and nitrogen species ; Figure 4 provides a simple illustration. The decreasing Fermi level that passes the antibonding LUMO of ammonia from the right will favor strong adsorption of ammonia molecules while the increasing Fermi level towards the antibonding LUMO of nitrogen from the left will favor desorption of neutral $ZN_{20}$ species.

Clearly, an alignment of the Fermi level of the catalyst between the antibonding LUMOs states of the ammonia and nitrogen adsorbates, $E_{N2}^{LUMO,a} < E_f < E_{NH3}^{LUMO,a}$, is optimal. Its positioning can be used to anticipate the rate limiting reaction step and dominating surface species; to the left of center results in limitation by nitrogen desorption, to the right of center results in limitation by ammonia adsorption. This result agrees with the classical qualitative Sabatier principle on the optimal binding energy and as well as quantitative theoretical results presented by Nørskov and coworkers [27].

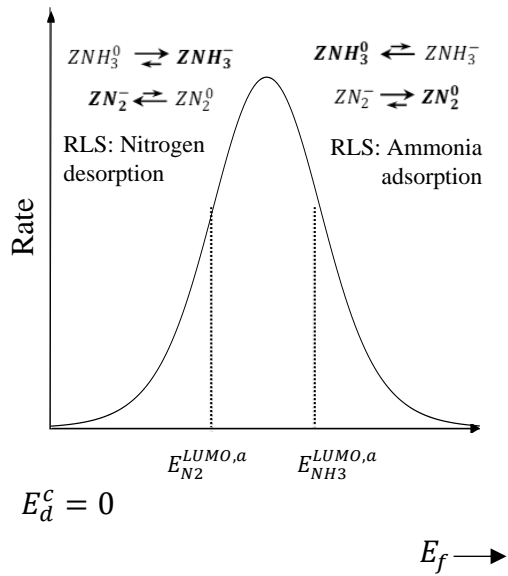

Figure 4: *Scheme for the ammonia decomposition reaction rate with respect to the Fermi level of an arbitrary metal catalyst ($E_d^c$ is set as a reference point =0 in our discussion). The antibonding LUMOs of adsorbed nitrogen and ammonia species are expected on either side of the rate maximum. RLS = rate limiting step.*

The optimal Fermi level of the catalyst can be determined where the reaction rate reaches its maximum:

$$E_f^{opt} = \frac{1}{2}\left(E_{NH3}^{LUMO,a} + E_{N2}^{LUMO,a}\right) + \frac{1}{2}(E_5 - E_2) + \frac{k_bT}{2}ln\left(\frac{1 + 1/K_4}{1 + 1/K_1C_{NH3}}\right). \qquad (9)$$

Where $E_2$ and $E_5$ are activation energies of steps (2) and (5) of the mechanism, $k_b$ represents Boltzmann's constant. It is seen from Eqn. (9) that when $\frac{1}{K_4} \approx \frac{1}{K_1C_{NH3}}$, the optimal Fermi level lies higher from the middle of antibonding states of the reagent (ammonia molecules) and the product (nitrogen molecules) $\left(E_{NH3}^{LUMO,a}, E_{N2}^{LUMO,a}\right)$ by a half the activation energy difference of steps 2 and 5. If chemisorption of ammonia molecules is strong due to the Fermi level of the catalyst located on the left of the antibonding state with ammonia (fig. 4), the fraction of negatively charged ammonia molecules will be large. However, when it is on the left of the antibonding state with nitrogen molecules, their neutralization and desorption is impaired. If chemisorption of ammonia molecules is weak due to the Fermi level of the catalyst lying on the right of the antibonding state with ammonia, then the fraction of negatively charged ammonia molecules will be small, although this situation is beneficial for neutralization of nitrogen molecules to be desorbed. Thus, there should be an optimal Fermi level relative to the antibonding LUMOs of ammonia and nitrogen molecules to ensure good chemisorption and desorption to maximize the reaction rate.

## 6. Reduced Mechanism: Nonlinear Analysis

To provide a rigorous solution to the reduced mechanism (Table 3), we can use a single route technique which is a simple conclusion of the theory of steady-state reactions by Horiuti-Temkin; as described in Yablonsky *et al.* [25]. The overall reaction equation results from addition of the elementary steps of the detailed mechanism by multiplying them by the number of times they occur in the catalytic cycle, which are called Horiuti numbers, Table 4.

*Table 4: Catalytic cycle, the Horuiti numbers, and overall reaction for ammonia decomposition.*

| Horiuti numbers | | |
|---|---|---|
| 2 | $NH_3 + Z \leftrightarrow ZNH_3^0$ | (4-1) |
| 2 | $ZNH_3^0 \leftrightarrow ZNH_3^-$ | (4-1*) |
| 2 | $2ZNH_3^- \rightarrow 2ZNH^-H_2^- + H_2$ | (4-2) |
| 1 | $2ZNH^- \rightarrow 2ZN^- + H_2$ | (4-3) |
| 1 | $2ZN^- \leftrightarrow ZN_2^- + Z$ | (4-4) |
| 1 | $ZN_2^- \leftrightarrow ZN_2^0$ | (4-4*) |
| 1 | $ZN_2^0 \rightarrow N_2 + Z$ | (4-5) |
| overall reaction | $NH_3 \rightarrow N_2 + 3H_2$ | |

The single route is characterized by the following relationships related to seven intermediates:

$$2R = R_1;\ 2R = R_{1*};\ 2R = R_2;\ R = R_3;\ R = R_4;\ R = R_{4*};\ R = R_5. \tag{10}$$

Here $R_1, R_{1*}, R_2, R_3, R_4, R_{4*}$ and $R_5$ are rates of corresponding steps, $R_1 = R_1^+ - R_1^-$; $R_{1*} = R_{1*}^+ - R_{1*}^-$; $R_4 = R_4^+ - R_4^-$; $R_{4*} = R_{4*}^+ - R_{4*}^-$, R is the route rate. In this case, we do not need to write equations for intermediates, and no assumptions are made about the abundance of empty active sites. The electronic steps (1*) and (4*) are assumed to be fast.

We can write the balance equation for the fractions of all intermediates:

$$[Z] + [ZNH_3^0] + [ZNH_3^-] + [ZNH^-] + [ZN^-] + [ZN_2^-] + [ZN_2^0] = 1. \tag{11}$$

The fractions of all seven intermediates can be expressed through the kinetic and the thermodynamics parameters and the route rate:

1. 
$$R_5 = R = k_5[ZN_2^0] \Rightarrow [ZN_2^0] = \frac{1}{k_5}R \qquad (12)$$

2. 
$$K_{4*} = \frac{[ZN_2^0]}{[ZN_2^-]}; \Rightarrow [ZN_2^-] = \frac{[ZN_2^0]}{K_{4*}} = \frac{1}{k_5 K_{4*}}R \qquad (13)$$

3. 
$$R_3 = R = k_3[ZNH^-]^2 \Rightarrow [ZNH^-] = \frac{1}{\sqrt{k_3}}\sqrt{R} \qquad (14)$$

4. 
$$R_2 = 2R = 2k_2[ZNH_3^-] \Rightarrow [ZNH_3^-] = \frac{2}{k_2}R \qquad (15)$$

5. 
$$K_{1*} = \frac{[ZNH_3^-]}{[ZNH_3^0]}; \Rightarrow [ZNH_3^0 N_2^-] = \frac{[ZNH_3^- N_2^0]}{K_{14*}} = \frac{12}{k_{25} K_{14*}}R \qquad (16)$$

6. 
$$R_1 = 2R = k_1^+ C_{NH3}[Z] - k_1^-[ZNH_3^0] = k_1^+ C_{N3}[Z] - k_1^- \frac{2}{k_2 K_{1*}} R \Rightarrow \qquad (17)$$
$$[Z] = 2R\left(\frac{1}{k_1^+ C_{NH3}} + \frac{1}{k_2}\frac{1}{C_{NH3} K_1 K_{1*}}\right)$$

7. 
$$R_4 = R = k_4^+[ZN^-]^2 - k_4^-[ZN_2^-][Z] \qquad (18)$$
$$[ZN^-]^2 = R\left(\frac{1}{k_4^+} + \frac{1}{k_5 K_{4*} K_4}[Z]\right) \Rightarrow$$
$$[ZN^-] = \sqrt{2\frac{1}{k_5 K_4 K_{4*}}\left(\frac{1}{k_1^+ C_{NH3}} + \frac{1}{k_2 K_1 C_{NH3} K_{1*}}\right) R^2 + \frac{1}{k_4^+}R}.$$

We can obtain a polynomial of the fourth power for the route rate by inserting Eqns. (12)-(18) into the balance Eqn. (11). We will not solve the quartic equation for the rate but instead we will analyze particular cases of the solution to reveal a volcano-shape dependence of the rate upon the Fermi level.

**Case 1**. $[Z] \approx 1$ (abundance of empty active sites). The route rate is described by the following formula:

$$\frac{1}{R} = \frac{2}{k_1^+ C_{NH3}} + \frac{2}{k_2}\frac{1}{C_{NH3} K_1 K_{1*}} \qquad (19)$$

In Eqn. (19), there is only one electronic term ($K_{1*}$), so the volcano-shape dependence cannot be observed.

**Case 2**. $[ZN_2^0]$ and $[ZN_2^-]$ are dominating, then we have:

$$\frac{1}{R} = \frac{1}{k_5}\left(1 + \frac{1}{K_{4*}}\right) \qquad (20)$$

It is seen that there is no volcano-behavior again.

**Case 3**. $[ZNH_3]$, $[ZN_2]$ and $[Z]$ are dominating:

$$\frac{1}{R} = \frac{1}{k_5} + \frac{1}{k_5}\frac{1}{K_{4*}} + \frac{2}{k_2} + \frac{2}{k_2}\frac{1}{K_{1*}} + \frac{2}{k_1^+ C_{NH3}} + \frac{2}{k_2}\frac{1}{C_{NH3} K_1 K_{1*}} \tag{21}$$

Here, there is an interplay between two electronic equilibrium constants $(K_{1*}, K_{4*})$, so in this case the volcano-dependence is observed. From this analysis, we can conclude that for an observation of a volcano-dependence it is necessary to have the simultaneous presence of intermediates of both reagents and products, and their amounts should be not small in comparison with empty sites on the catalyst.

### 7. Discussion

The essential feature of Wolkenstein's theory is the localization of free charge carriers (electrons or holes) onto the chemisorbed species that results in a change in the nature of the chemical bond with the surface. As a result, charged species will exhibit different reactivity compared with the same neutral species. A fast equilibrium exists for transition between charged and neutral and the relative surface population of the two species depends on the Fermi level of the catalyst. It is worth noting the definition made by Ilya Lifshitz, who considered a metal as "a solid with a Fermi surface". He expressed that it determined nearly all of a metal's properties. Lifshitz wrote, "A knowledge of the Fermi surface where filled and empty states are adjacent, enables one to understand most of the mechanical, electrical, and magnetic properties of a metal" [28]. According to Wolkenstein's electronic theory, a number of characteristics of chemical reactions catalyzed by semiconductors depend on the Fermi level of the catalyst:

1. the relative number of different forms of chemisorption, distinguished by the character of the chemical bond of adsorbed molecule with the surface;
2. the number of chemisorbed molecules at equilibrium with the gas phase at a given pressure and temperature;
3. the catalytic activity of the surface with respect to a given reaction;
4. the selectivity of a catalyst with respect to two (or more) concurrently proceeding reactions.

Comparison of the results of the linear analysis, Eqn.(6), with the non-linear analysis, Eqn.(21), demonstrates these common features:

- in both cases, there is an interplay between two sets of steps, ammonia adsorption-dehydrogenation steps (1,1*,2) which includes the electronic step (1*), and nitrogen desorption steps (4*,5) which includes the electronic step (4*);
- in both cases, functional dependences of 1/R are identical qualitatively.

Both eqns. (6) and (21) indicate the dependence of the rate of ammonia decomposition as a function of the Fermi level of the catalyst. A decreasing Fermi level towards the antibonding state with ammonia is preferable for the adsorption of ammonia molecules to increase the amount of negatively charged (strongly bound) $ZNH_3^-$. An increasing Fermi level towards the antibonding state with nitrogen is preferable for desorption of nitrogen molecules from the surface of the catalyst.

*Volcano-shaped Rate Dependence*

The mechanism of ammonia decomposition was previously investigated over different catalyst [8] and we will show that our model agrees with the relationship of different metals found therein and reproduced in Figure 1 which plots the ammonia decomposition rate as a function of the dissociative nitrogen adsorption energy. Recall from Eqn. (7) that the thermodynamic parameters of electronic steps are dictated by a competition of exponential terms for the Fermi level of catalyst and the antibonding LUMO of the adsorbed species. Generally, the Fermi level, or experimental work function, $\phi$, can be measured. The antibonding LUMO of adsorbed species might be calculated but such information is not widely available for different adsorbates. This information can be gleaned from the location of the metal *d*-band center which is readily measured or calculated. Available data for metals relevant to ammonia decomposition are collected in Table 5. The work function values are listed for reference to the values of gas phase molecular orbitals described in section 3; recall the work function relation to the Fermi level: $e\phi = E_{vac} - E_f$, where $e$ is the charge of an electron and $E_{vac}$ the vacuum potential. The location of the Fermi level relative to the *d*-band center was calculated by Hammer and Nørskov [2].

*Table 5: Work function and relative* d-*band center reported for different metals.*

| Metal | Work function, $\phi$, (eV, relative to vacuum level) [29] | Fermi Level (eV, relative to *d*-band center) [2] | *d*-coupling matrix element squared, $V_{ad}^2$ [2] |
| --- | --- | --- | --- |
| Fe | 4.5 | 0.92 | 1.59 |
| Ru | 4.71 | 1.41 | 3.87 |
| Co | 5.0 | 1.17 | 1.34 |
| Ni | 5.15 | 1.29 | 1.16 |

If we fix the *d*-band center at zero for reference, then we can compare materials based on the reported distance of the Fermi level. Let us first compare data for iron and cobalt where the Fermi level of iron is located closer to the *d*-band center. This means that for iron, the Fermi level is closer to $E_{N2}^{LUMO,a}$ and further from $E_{NH3}^{LUMO,a}$ while for cobalt the Fermi level is closer to $E_{NH3}^{LUMO,a}$ and further from $E_{N2}^{LUMO,a}$. So, iron can be considered to the left of the maximum in Figure 4 and cobalt, to the right. In terms of Eqn. (7), these relative separations result in a higher population of the both negatively charged *ZNH3-* and *ZN2-* species on iron; more favorable adsorption of NH3 and a decrease in the facility for N2 desorption. For cobalt, according to the terms of Eqn. (7), these relative separations can also be described as a shift of the equilibrium of electronic steps (3-1*) and (3-4*) away from the charged species, leading to more neutrally charged species *ZNH3$^0$* and *ZN2$^0$* and thus, less favorable adsorption of NH3 but facilitated desorption of N2. The relative location of the Fermi level changes the rate according to the two competing thermodynamic terms, $K_{1*}$ and $K_{4*}$ in Eqn. (6) and the placement of iron and cobalt in terms of electronic equilibrium argument agrees with that of Figure 1.

Now, if we next consider nickel, according to Table 5 this should sit further to the right of cobalt in Figure 4. Likewise, ruthenium should be even further. The placement of nickel agrees with Figure 1 but ruthenium is found near the maximum. This inconsistency is explained by a higher location of the antibonding LUMOs of the adsorbate due to the larger degree of adsorbate-metal orbital overlap according to the *d*-coupling matrix element $V_{ad}^2$, Table 5 [2]; *i.e.* greater *d*-band splitting compared to that of cobalt or nickel. The locations of $E_{N2}^{LUMO,a}$ and $E_{NH3}^{LUMO,a}$ in Figure 4 would be significantly different for ruthenium (Row 5), compared to iron, cobalt and nickel (Row 4). Relative comparison of the Fermi level across the rows is more reliable then moving downwards through the periodic table where the *d*-coupling matrix element changes more significantly. This illustrates that it is the interplay of the Fermi level, the *d*-band center and the location of the antibonding LUMOs together dictate the relative population of charged and neutral species on the surface and hence, the direction of the reaction mechanism.

Recently, the mechanism of ammonia decomposition was investigated over bimetallic catalysts, for example in a bimetallic CoFe under both steady state and transient conditions [6]. In the steady-state reaction, the rate of ammonia transformation of the CoFe bimetallic was found to be greater than that of either constituent; a volcano-type dependence. As for the CoFe bimetallic, according to DFT calculations, the Fermi level of bimetallic materials with different compositions can be predicted as a linear function of the surface coverage of one component on the surface of the second component [30]:

$$E_{f,1-2} = x_{s,1}E_{f,1} + x_{s,2}E_{f,2} \tag{22}$$

where $x_{s,I}$ is the surface area ratio of the $i_{th}$ component. According to Eqn. (22) the Fermi level of CoFe should be between the Fermi levels of iron and cobalt depending on the surface coverage of cobalt on iron. From another side, based on the experimental data in Pd-shell–Pt-core nanoparticles [31], we can suppose that in core-shell bimetallic catalysts as a result of tensile strain in the Co overlayer deposited on the iron core, the Fermi level shifts towards the *d*-band center compared to that for the cobalt catalyst, and the energy difference between the Fermi level and the antibonding LUMOs of the adsorbates actually become smaller for the bimetal. Compared to the cobalt catalyst, this shifts the equilibrium of electronic steps back towards the charged species and improves the adsorption of ammonia while nitrogen desorption still occurs with ease, and the reaction rate in Eqn. (6) changes accordingly. The steady-state experimental result of Wang *et al.* agrees with this conclusion [6]. Thus, optimization of the reaction rate is achieved by manipulation of the Fermi level through alloying of metals with respect to the *d*-band center. Alternatively, other works have described how the Fermi level/work function can be varied according to the size of nanoparticle domains [32, 33] as well as the addition of alkali promoters [34, 35].

## 8. Conclusion

Starting with the traditional kinetic model of ammonia decomposition, we have added explicit steps describing the participation of free electrons from the catalyst through the LUMOs of surface adsorbates acting as a conduit to create charged surface species. This is based on the electronic theory of catalysis proposed by Th. Wolkenstein for semiconductors. The Fermi level was used to determine the availability of free charge carriers that can change the nature of the adsorbate. We have extended the Wolkenstein method by application of *d*-band theory that describes the splitting of adsorbate states into bonding and antibonding. Via the adsorbate LUMOs, catalyst electrons may localize and change the nature of the surface species. The relative population of charged versus neutral adsorbates depends on the location of the Fermi level of the catalyst and the extent of *d*-band splitting. Charged and neutral surface intermediates will exhibit different reactivity and hence the reaction rate also depends on the Fermi level. Based on the electronic kinetic model, we have presented a rate expression that demonstrates a volcano-shaped dependence on the Fermi level. In difference from the classical Sabatier volcano-curve, considering a single reactant, the volcano-shaped reaction rate of the presented model is obtained via the interplay of both reactant adsorption and product desorption, which is achieved by considering both electronic and *d*-band theory based on participation of charged species in the kinetic model. The locations of atomic and molecular orbitals for intermediates with respect to the Fermi level of the catalyst are crucial for better understanding of the detailed reaction mechanism and the influence of electronic participation of the catalyst. In this sense, the knowledge of the molecular orbitals for intermediates is a challenge for DFT calculations and spectroscopic experiments.

**Acknowledgements**
RF acknowledges support from the U.S. Department of Energy (USDOE), Office of Energy Efficiency and Renewable Energy (EERE), Advanced Manufacturing Office Next Generation R&D Projects under contract no. DE-AC07-05ID14517.